\documentclass{llncs}
\usepackage{a4,a4wide}
\usepackage{epsfig}
\usepackage{makeidx}		

\newif\ifREWRITING
\REWRITINGtrue
\REWRITINGfalse

\newlength{\layoutwidth}
\newlength{\halffigwidth}
\newlength{\smallfigwidth}
\newlength{\largefigwidth}

\ifREWRITING
\fi
\newcommand{\LW}[1]{\smash{\lower2.ex\hbox{#1}}}

\newcommand{\Code}[1]{{\tt #1}}

\begin{document}

\mainmatter			

\title{Translating Nondeterministic Functional Language based on
  Attribute Grammars into Java}

\author{
Masanobu Umeda\inst{1} \and
Ryoto Naruse\inst{2} \and
Hiroaki Sone\inst{2} \and
Keiichi Katamine\inst{1}
}

\institute{
Kyushu Institute of Technology, 680-4 Kawazu, Iizuka 820-8502, Japan\\
\email{umerin@ci.kyutech.ac.jp}
\and
NaU Data Institute Inc., 680-41 Kawazu, Iizuka 820-8502, Japan\\
}

\maketitle

\begin{abstract}
Knowledge-based systems are suitable for realizing advanced functions
that require domain-specific expert knowledge, while knowledge
representation languages and their supporting environments are
essential for realizing such systems.  Although Prolog is useful and
effective in realizing such a supporting environment, the language
interoperability with other implementation languages, such as Java, is
often an important issue in practical application development.  This
paper describes the techniques for translating a knowledge
representation language that is a nondeterministic functional language
based on attribute grammars into Java.  The translation is based on
binarization and the techniques proposed for Prolog to Java
translation although the semantics are different from those of Prolog.
A continuation unit is introduced to handle continuation efficiently,
while the variable and register management on backtracking is
simplified by using the single and unidirectional assignment features
of variables.  An experimental translator written in the language
itself successfully generates Java code, while experimental results
show that the generated code is over 25 times faster than that of
Prolog Cafe for nondeterministic programs, and over 2 times faster for
deterministic programs.  The generated code is also over 2 times
faster than B-Prolog for nondeterministic programs.
\end{abstract}

\section{Introduction}\label{introduction}

There is high demand for advanced information services in various
application domains such as medical services and supply-chain
management, as information and communication technology penetrates
deeply into our society.  Clinical decision support
\cite{Kaplan2001a:CDS,Takada2007a:CAFE} to prevent medical errors and
order placement support for optimal inventory management
\cite{Nagasawa2006a:LOGISTICS} are typical examples.  It is, however,
not prudent to implement such functions as a normal part of the
traditional information system using conventional programming
languages.  This is because expert knowledge is often large scale and
complicated, and each application domain typically has its own
specific structures and semantics.  Therefore, not only the analysis,
but also the description, audit, and maintenance of such knowledge are
often difficult without expertise in the application domain.  It is
thus, essential to realize such advanced functions to allow domain
experts themselves to describe, audit, and maintain their knowledge.
A knowledge-based system approach is suitable for such purposes
because a suitable framework for representing and managing expert
knowledge is supplied.

Previously, Nagasawa et al. proposed the knowledge representation
language DSP \cite{Nagasawa1990a:DSP,Umeda1996a:DSP} and its
supporting environment.  DSP is a nondeterministic functional language
based on attribute grammars
\cite{Katayama1983a:AttributeGrammars,LNCS461} and is suitable for
representing complex search problems without relying on any side
effects.  The supporting environment has been developed on top of an
integrated development environment called Inside Prolog
\cite{Katamine2004a:INSIDE}.  Inside Prolog provides standard Prolog
functionality, conforming to ISO/IEC 13211-1 \cite{ISO1995a:PROLOG},
and also a large variety of Application Programming Interfaces (APIs)
that are essential for practical application development and
multi-thread capability for enterprise use \cite{Umeda2006a:INSIDE}.

These features allow the consistent development of knowledge-based
systems from prototypes to practical systems for both stand-alone and
enterprise use \cite{Umeda2007a:INSIDE}.  Such systems have been
applied to several practical applications, and the effectiveness
thereof has been clarified.  However, several issues have also been
perceived from these experiences.  One is the complexity of combining
a Prolog-based system with a system written in a normal procedural
language, such as Java.  The other is the adaptability to a new
computer environment such as mobile devices.

This paper describes the implementation techniques required to
translate a nondeterministic functional language based on attribute
grammars into a procedural language such as Java.  The proposed
techniques are based on the techniques for Prolog to Java translation.
Section 2 gives an overview of the knowledge representation language
DSP, and clarifies how it differs from Prolog.  In Section 3, the
translation techniques for logic programming languages are briefly
reviewed, and basic ideas useful for the translation of DSP
identified.  Section 4 discusses the program representations of DSP in
Java, while Section 5 evaluates the performance using an experimental
translator.

\section{Overview of Knowledge Representation Language DSP}\label{overview}
\subsection{Background}

It is essential to formally analyze, systematize, and describe the
knowledge of an application domain in the development of a
knowledge-based system.  The description of knowledge is conceptually
possible in any conventional programming language.  Nevertheless, it
is difficult to describe, audit, and maintain a knowledge base using a
procedural language such as Java.  This is because the knowledge of an
application domain is often large scale and complicated, and each
application domain has its own specific structures and semantics.  In
particular, the audit and maintenance of written knowledge is a major
issue in an information system involving expert knowledge, because
such a system is very often stiffened and the transfer of expert
knowledge to succeeding generations is difficult
\cite{Nagasawa1988a:ICAD}.  Therefore, it is very important to provide
a framework to enable domain experts themselves to describe, audit,
and maintain their knowledge included in an information system
\cite{Umeda2009a:FORGING}.  It is perceived that a description
language that is specific to an application domain and is designed so
as to be described by domain experts is superior in terms of the
minimality, constructibility, comprehensibility, extensibility, and
formality of the language \cite{Hirota1995a:MODELING}.  For this
reason, Prolog cannot be considered as a candidate for a knowledge
representation language.

DSP is a knowledge representation language based on nondeterministic
attribute grammars.  It is a functional language with a search
capability using the generate and test method.  Because the language
is capable of representing trial and error without any side-effects or
loop constructs, and the knowledge descriptions can be declaratively
read and understood, it is suitable for representing domain-specific
expert knowledge involving search problems.

\subsection{Syntax and Semantics of DSP}

A program unit to represent knowledge in DSP is called a ``module'',
and it represents a nondeterministic function involving no
side-effects.  Inherited attributes, synthesized attributes, and
tentative variables for the convenience of program description, all of
which are called variables, follow the single assignment rule and the
assignment is unidirectional.  Therefore, the computation process of a
module can be represented as non-cyclic dependencies between
variables.

\begin{table}[htb]
\begin{center}
\caption{Typical statements in the DSP language}
\begin{tabular}{l|l|p{8cm}} \hline \hline
\multicolumn{1}{c|}{Type} &
\multicolumn{1}{|c|}{Statement} &
\multicolumn{1}{|c}{Function} \\ \hline
generator & \Code{for(B,E,S)} & 
Assume a numeric value from B to E with step S \\
generator & \Code{select(L)} &
Assume one of the elements of a list L \\
generator & \Code{call(M,I,O)} & 
Call a module M nondeterministically with inputs I and outputs O \\
calculator & \Code{dcall(M,I,O)} & 
Call a module M deterministically with inputs I and outputs O \\
calculator & \Code{find(M,I,OL)} &
Get a list OL of all outputs of a module M with inputs I \\
tester & \Code{when(C)} & Specify the domain C of a method \\
tester & \Code{test(C)} & Specify the constraint C of a method \\
tester & \Code{verify(C)} & Specify the verification condition C \\
\hline
\end{tabular}
\label{statement-example}
\end{center}
\end{table}

Table \ref{statement-example} shows some typical statements in the
language.  In this table, the types, generator, calculator, and
tester, are functional classifications in the generate and test
method.  Generators \Code{for(B,E,S)} and \Code{select(L)} are
provided as primitives for the convenience of knowledge representation
although they can be defined as modules using the nondeterministic
features of the language.  Both \Code{call(M,I,O)} and
\Code{dcall(M,I,O)} are used for module decomposition, with the latter
restricting the first solution of a module call like \Code{once/1} in
Prolog \footnote{\Code{dcall} stands for deterministic call.}, while
the former calls a module nondeterministically.  Calculator
\Code{find(M,I,OL)} collects all outputs of a module and returns a
list thereof.  Testers \Code{when(C)} and \Code{test(C)} are used to
represent decomposition conditions.  Both behaves in the same way in
normal execution mode \footnote{Failures of \Code{when(C)} and
  \Code{test(C)} are treated differently in debugging mode because of
  their semantic differences.}, although the former is intended to
describe a guard of a method, while the latter describes a constraint.
Tester \Code{verify(C)} does not affect the execution of a module
although it is classified as the tester.  Solutions in which a
verification condition is not satisfied are indicated as such, and
these verification statuses are used to evaluate the inference
results.

\begin{figure}[hbt]
\begin{center}
\begin{verbatim}
    pointInQuarterCircle({R : real},           --(a)
                         {X : real, Y : real}) --(b)
      method
        X : real = for(0.0, R, 1.0);           --(c)
        Y : real = for(0.0, R, 1.0);           --(d)
        D : real = sqrt(X^2 + Y^2);            --(e)
        test(D =< R);                          --(f)
      end method;
    end module;
\end{verbatim}
\caption{Module \Code{pointInQuarterCircle}, which enumerates all points
  in a quarter circle}
\label{module-example}
\end{center}
\end{figure}

\begin{figure}[hbt]
\begin{center}
\epsfig{file=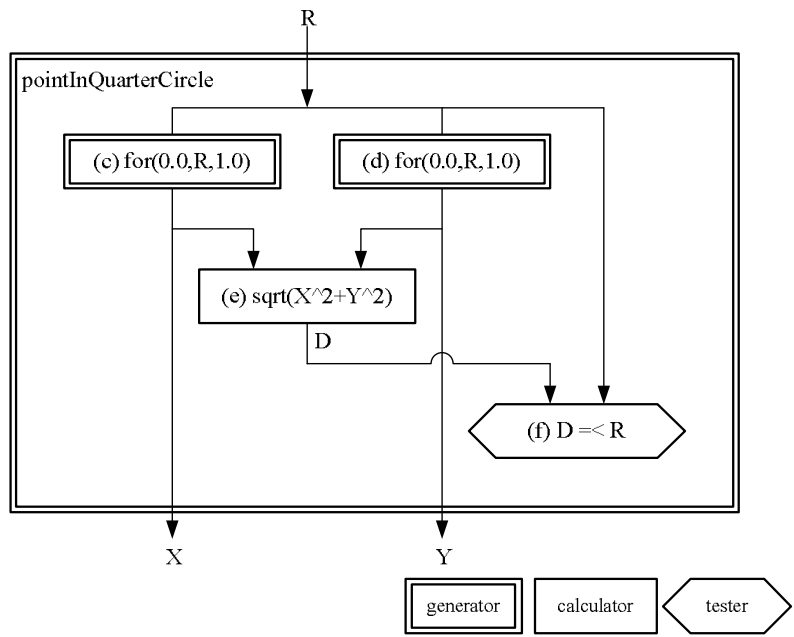,width=\smallfigwidth}
\caption{Data flow diagram of module \Code{pointInQuarterCircle}}
\label{data-flow-example}
\end{center}
\end{figure}

Figure \ref{module-example} gives the code for module
\Code{pointInQuarterCircle}, which enumerates all points in a quarter
circle with radius \Code{R}.  Statements (a) and (b) in
Fig. \ref{module-example} define the input and output variables of
module \Code{pointInQuarterCircle}, respectively.  Statements (c) and
(d) assume the values of the variables \Code{X} and \Code{Y} from
\Code{1} to \Code{R} with an incremental step \Code{1}.  Statement (e)
calculates the distance \Code{D} between point \Code{(0,0)} and point
\Code{(X,Y)}.  Statement (f) checks if point \Code{(X,Y)} is within
the circle of radius \Code{R}.  Module \Code{pointInQuarterCircle}
runs nondeterministically for a given \Code{R}, and returns one of all
possible \Code{\{X,Y\}} values
\footnote{\Code{\{X,Y\}} represents a vector of two elements \Code{X}
  and \Code{Y}.}.  Therefore, this module also behaves as a generator.
Statements (c) to (f) can be listed in any order, and they are
executed according to the dependencies between variables.  Therefore,
the computation process can be described as a non-cyclic data flow.
Figure \ref{data-flow-example} shows the data flow diagram for module
\Code{pointInQuarterCircle}.  Because no module includes any
side-effects, the set of points returned by the module for the same
input is always the same.

Figure \ref{for-example} shows an example of module \Code{for}, which
implements the generator primitive \Code{for}.  If multiple methods
are defined in a module with some overlap in their domains specified
by \Code{when}, the module works nondeterministically, and thus a
module can also be a generator.  In this example, there is overlap
between the domains specified by statements \Code{(a)} and \Code{(c)}.

\begin{figure}[hbt]
\begin{center}
\begin{verbatim}
    for({B : real, E : real, S : real},{N : real})
      method                        --The fist method
        when(B =< E);               --(a)
        N : real = B;               --(b)
      end method;
      method                        --The second method
        when(B+S =< E);             --(c)
        B1 : real = B+S;            --(d)
        call(for, {B1, E, S}, {N}); --(e)
      end method;
    end;
\end{verbatim}
\caption{Module \Code{for}, which implements the generator primitive \Code{for}}
\label{for-example}
\end{center}
\end{figure}

\subsection{Execution Model for DSP}

Since the variables follow the single assignment rule and the
assignment is unidirectional, the statements are partially ordered
according to the dependencies between variables.  In the execution,
the statements must be totally reordered and evaluated in this order.
Although the method used to order the partially ordered statements
totally does not affect the set of solutions, the order of the
generators affects the order of the solutions returned from a
nondeterministic module.

The execution model for DSP can be represented in Prolog.  Figure
\ref{execution-model} illustrates an example of a simplified DSP
interpreter in Prolog.  In this interpreter, statements are
represented as terms concatenated by ``;'' and it is assumed that the
statements are totally ordered.  Variables are represented using
logical variables in Prolog.  Actually, the development environment
for DSP provides a compiler that translates into Prolog code, with the
generated Prolog code translated into bytecode by the Prolog compiler
in the runtime environment.

\begin{figure}[hbt]
\begin{center}
\begin{verbatim}
    solve((A ; B)) :-
        solve(A),
        solve(B).
    solve(call(M, In, Out)) :-
        reduce(call(M, In, Out), Body),
        solve(Body).
    solve(dcall(M, In, Out)) :-
        reduce(call(M, In, Out), Body),
        solve(Body),!.
    solve(find(M, In, OutList)) :-
        findall(Out, solve(M, In, Out), OutList).
    solve(when(Exp)) :-
        call(Exp),!.
    solve(test(Exp)) :-
        call(Exp),!.
    solve(V := for(B, E, S)) :- !,
        for(B, E, S, V).
    solve(V := select(L)) :- !,
        member(V, L).
    solve(V := Exp) :-
        V is Exp.
\end{verbatim}
\caption{Simplified DSP interpreter in Prolog}
\label{execution-model}
\end{center}
\end{figure}

\section{Translation Techniques for Logic Programming Languages}

Prolog is a logic programming language that offers both declarative
features and practical applicability to various application domains.
Many implementation techniques for Prolog and its family have been
proposed, while abstract machines represented by the WAM (Warren's
Abstract Machine) \cite{AIT1991a:PROLOG} have proven effective
practical implementation techniques.  On the other hand, few Prolog
implementations provide practical functionality applicable to both
stand-alone systems and enterprise-mission-critical information
systems without using other languages.  Practically, Prolog is often
combined with a conventional procedural language, such as Java, C, and
C\#, for use in practical applications.  In such cases, language
interoperability is an important issue.

Language translation is one possible solution for improving the
interoperability between Prolog and other combined languages.  jProlog
\cite{Demoen1996:PROLOG} and Prolog Cafe \cite{Banbara2005a:PROLOG}
are Prolog to Java translators based on binarization
\cite{Tarau1990:PROLOG}, while P\# \cite{Cook2004a:PROLOG} is a Prolog
to C\# translator based on Prolog Cafe with concurrent extensions.
The binarization with continuation passing is a useful idea for
handling nondeterminism simply in procedural languages.  For example,
the following clauses
\begin{verbatim}
    p(X) :- q(X, Y), r(Y).
    q(X, X).
    r(X).
\end{verbatim}
can be represented by semantically equivalent clauses that take a
continuation goal \Code{Cont} as the last parameter:
\begin{verbatim}
    p(X, Cont) :- q(X, Y, r(Y, Cont)).
    q(X, X, Cont) :- call(Cont).
    r(X, Cont) :- call(Cont).
\end{verbatim}
Once clauses have been transformed into this form, clauses composing a
predicate can be translated into Java classes.  Figure
\ref{PrologCafe-example1} gives an example of code generated by Prolog
Cafe.  Predicate \Code{p/2} after binarization is represented as a
Java class called \Code{PRED\_p\_1}, which is a subclass of class
\Code{Predicate}.  The parameters of a predicate call are passed as
the arguments of the constructor of a class, while the right hand side
of a clause is expanded as method \Code{exec}.

\begin{figure}[hbt]
\begin{center}
\begin{verbatim}
    public class PRED_p_1 extends Predicate {
        public Term arg1;
    
        public PRED_p_1(Term a1, Predicate cont) {
            arg1 = a1;
            this.cont = cont; /* this.cont is inherited. */
        }
        ...
        public Predicate exec(Prolog engine) {
            engine.setB0();
            Term a1, a2;
            Predicate p1;
            a1 = arg1;
            a2 = new VariableTerm(engine);
            p1 = new PRED_r_1(a2, cont);
            return new PRED_q_2(a1, a2, p1);
        }
    }
\end{verbatim}
\caption{Java code generated by Prolog Cafe}
\label{PrologCafe-example1}
\end{center}
\end{figure}

If a predicate consists of multiple clauses as in the following
predicate \Code{p/1}, it may have choice points.
\begin{verbatim}
    p(X) :- q(X, Y), r(Y).
    p(X) :- r(X).
\end{verbatim}
In such a case, the generated code becomes more complex than before
because the choice points of \Code{p/1} must be handled for
backtracking.  Figure \ref{PrologCafe-example2} gives an example of
the generated code for predicate \Code{p/1} in the previous example.
Each clause of a predicate is mapped to a subclass of a class
representing the predicate.  In this example, classes
\Code{PRED\_p\_1\_1} and \Code{PRED\_p\_1\_2} correspond to the two
clauses of predicate \Code{p/1}.  Methods \Code{jtry} and \Code{trust}
of the Prolog engine correspond to WAM instructions that manipulate
stacks and choice points for backtracking.  The key ideas in Prolog
Cafe are that continuation is represented as an instance of a Java
class representing a predicate, and the execution control including
backtracking follows the WAM.  The translation is straightforward
through the WAM, while the interoperability with Java-based systems is
somewhat improved.  On the other hand, the disadvantage is the
performance of the generated code.

\begin{figure}
\begin{center}
\small
\begin{verbatim}
    public class PRED_p_1 extends Predicate {
        static Predicate _p_1_sub_1 = new PRED_p_1_sub_1();
        static Predicate _p_1_1 = new PRED_p_1_1();
        static Predicate _p_1_2 = new PRED_p_1_2();
        public Term arg1;

        ...
        public Predicate exec(Prolog engine) {
            engine.aregs[1] = arg1;
            engine.cont = cont;
            engine.setB0();
            return engine.jtry(_p_1_1, _p_1_sub_1);
        }
    }

    class PRED_p_1_sub_1 extends PRED_p_1 {
        public Predicate exec(Prolog engine) {
            return engine.trust(_p_1_2);
        }
    }

    class PRED_p_1_1 extends PRED_p_1 {
        public Predicate exec(Prolog engine) {
            Term a1, a2;
            Predicate p1;
            Predicate cont;
            a1 = engine.aregs[1];
            cont = engine.cont;
            a2 = new VariableTerm(engine);
            p1 = new PRED_r_1(a2, cont);
            return new PRED_q_2(a1, a2, p1);
        }
    }

    class PRED_p_1_2 extends PRED_p_1 {
        public Predicate exec(Prolog engine) {
            Term a1;
            Predicate cont;
            a1 = engine.aregs[1];
            cont = engine.cont;
            return new PRED_r_1(a1, cont);
        }
    }
\end{verbatim}
\caption{Java code with choice points generated by Prolog Cafe}
\label{PrologCafe-example2}
\end{center}
\end{figure}

\section{Program Representation in Java and Inference Engine}\label{representation}

This section describes the translation techniques for the
nondeterministic functional language DSP into Java based on the
translation techniques for Prolog.  Current implementations of the
compiler and inference engine for DSP have been developed on top of
Inside Prolog with the compiler generating Prolog code.  Therefore, it
is possible to translate this generated Prolog code into Java using
Prolog Cafe.  However, there are several differences between DSP and
Prolog in terms of the semantics of variables and the determinism of
statements.  These differences allow several optimizations in
performance, and the generated code can run faster than the code
generated by Prolog Cafe for compatible Prolog programs.  Fundamental
ideas of our translation techniques utilize the single and
unidirectional assignment features of variables and the deterministic
features of some statements.

The overall structure of the Java code translated from DSP provides
for one module being mapped to a single Java class, and each method in
a module mapped to a single inner class of the class.  Figure
\ref{PointInQuarterCircle-example1} shows an example of Java code for
module \Code{pointInQuarterCircle} given in Fig. \ref{module-example}.
Inner classes are used to represent an execution context of a
predicate as an internal state of a class instance.  Therefore, the
instances of an inner class are not declared as static unlike classes
in Fig. \ref{PrologCafe-example2}.

An overview of the translation process follows.  First, the data flow
of a module is analyzed for each method based on the dependencies
between variables, and the statements are reordered according to the
analysis results.  Next, the statements are grouped into translation
units called continuation units, and Java code is generated for each
method according to the continuation units.

\subsection{Data Flow Analysis}

As described in Sect. \ref{overview}, it is necessary to reorder and
evaluate statements so as to fulfill variable dependencies since
statements can be listed in any order.  Therefore, partially ordered
statements must first be totally reordered.  In the reordering
process, the order of the generators should be kept as long as the
variable dependencies are satisfied, because the order of generators
affects the order of the solutions as described in
Sec. \ref{overview}.  On the other hand, calculators or testers can be
moved forward for the least commitment as long as partial orders are
kept.

\subsection{Continuation Unit}

If statements of a method are totally ordered, they can be divided
into several groups of statements.  Each group is called a
continuation unit and consists of a series of deterministic
statements, such as calculators and testers, followed by a single
generator.  It should be noted that a continuation unit may not
contain a generator if it is the last one in a method.  In the
translation, a continuation unit is treated as a unit to translate,
and is mapped to a Java class representing a continuation.

In the example in Fig. \ref{PointInQuarterCircle-example1}, module
\Code{pointInQuarterCircle} has one method, and there are three
continuation units in the method.  Inner class \Code{Method\_1}
corresponds to this method of the module, and class
\Code{Method\_1\_cu1} corresponds to the continuation unit for
statement (c), class \Code{Method\_1\_cu2} to one for statement (d),
and class \Code{Method\_1\_cu3} to one for statements (e) and (f),
respectively.

\begin{figure}
\begin{center}
\small
\begin{verbatim}
    public class PointInQuarterCircle implements Executable {
      private Double r;
      private Variable x;
      private Variable y;
      private Executable cont;
      public PointInQuarterCircle(Double r,
                                  Variable x, Variable y, Executable cont)
      {
        this.r = r;
        this.x = x;
        this.y = y;
        this.cont = cont;
      }

      public Executable exec(VM vm) {
        return (new Method_1()).exec(vm);
      }

      public class Method_1 implements Executable {
        private Variable d = new Variable();
        private Executable method_1_cu1 = new Method_1_cu1();
        private Executable method_1_cu2 = new Method_1_cu2();
        private Executable method_1_cu3 = new Method_1_cu3();
    
        public Executable exec(VM vm) {
          return method_1_cu1.exec(vm);
        }

        class Method_1_cu1 implements Executable {
          public Executable exec(VM vm) {
            return new ForDouble(0.0, r.doubleValue(), 1.0, x, method_1_cu2);
          }
        }
        
        class Method_1_cu2 implements Executable {
          public Executable exec(VM vm) {
            return new ForDouble(0.0, r.doubleValue(), 1.0, y, method_1_cu3);
          }
        }

        class Method_1_cu3 implements Executable {
          public Executable exec(VM vm) {
            d.setValue(Math.sqrt(x.doubleValue()*x.doubleValue() +
                                 y.doubleValue()*y.doubleValue()));
            if(!(d.doubleValue() <= r.doubleValue())){
              return Executable.failure;
            }
            return cont;
          }
        }
      }
    }
\end{verbatim}
\caption{Java code generated for module \Code{pointInQuarterCircle}}
\label{PointInQuarterCircle-example1}
\end{center}
\end{figure}

\subsection{Variable and Parameter Passing}

Although variables follow the single assignment rule like Prolog, the
binding of a variable is unidirectional unlike Prolog.  Therefore, it
is not necessary to introduce logical variables and unification,
unlike in Prolog Cafe.  This also implies that the trail stack and
variable unbinding using the stack are unnecessary on backtracking.
Therefore, a class representing the variables is only necessary as a
place holder for the output values of a module.  Class \Code{Variable}
is introduced to represent such variables.

Prolog Cafe uses the registers of the Prolog VM to manage the
arguments of a goal.  This approach is consistent with the WAM, but is
sometimes inefficient since it requires arguments to be copied from/to
registers to/from the stack on calls and backtracking.  On the other
hand, because the direction of variable binding is clearly defined in
DSP, it is unnecessary to restore variable bindings on backtracking as
described before.  Instead, variables can always be overwritten when a
goal is re-executed after backtracking.  Therefore, input and output
parameters can be passed as arguments of a class constructor.  This
simplifies the management of variables and arguments.  In addition, as
shown in Fig. \ref{PointInQuarterCircle-example1}, basic Java types,
such as \Code{int} and \Code{double}, can be passed directly as inputs
in some cases.  This contributes to the performance improvement.

\subsection{Inference Engine}

An inference engine for the translated code is very simple because
management of variables and registers on backtracking is unnecessary.
Figure \ref{engine-example} shows an example of the inference engine
called \Code{VM}, which uses a stack represented as an array of
interface \Code{Executable} to store choice points.  Method
\Code{call()} is an entry point to call the module to find an initial
solution, while method \Code{redo()} is used to find the next
solution.  A typical call procedure of a client program in Java is
given below.
\begin{verbatim}
    VM vm = new VM();
    Double r = new Double(10.0);
    Variable x = new Variable();
    Variable y = new Variable();
    Executable m = new PointInQuarterCircle(r, x, y,
                                            Executable.success);
    for (boolean s = vm.call(m); s == true; s = vm.redo()) {
      System.out.println("X=" + x.doubleValue() +
                         ", Y=" + y.doubleValue());
    }
\end{verbatim}
This client program creates an inference engine, prepares output
variables to receive the values of a solution, creates an instance of
class \Code{PointInQuarterCircle} with inputs and outputs, and calls
\Code{call()} to find an initial solution.  It then calls
\Code{redo()} to find the next one until there are no more solutions.

Because the implementation of the inference engine is simple and
multi-thread safe, and the generated classes of a module are also
multi-thread safe, it is easy to deploy instances of the engine in a
multi-thread environment.

\begin{figure}[hbt]
\begin{center}
\small
\begin{verbatim}
    public class VM {
      private Executable[] choicepoint;
      private int ccp = -1; // Current choice point.
      ...
    
      public VM(int initSize) {
        choicepoint = new Executable[initSize];
      }
      ...
      public boolean call(Executable goal) {
        while (goal != null) {
          goal = goal.exec(this);
          if (goal == Executable.success) {
            return true;
          } else if (goal == Executable.failure) {
            goal = getChoicePoint();
          }
        }
        return false;
      }
    
      public boolean redo() {
        return call(getChoicePoint());
      }
    }
\end{verbatim}
\caption{Inference engine for DSP}
\label{engine-example}
\end{center}
\end{figure}

\section{Implementation and Performance Evaluation}

We have implemented the translator for DSP into Java based on the
techniques proposed in Sec. \ref{representation}.  The translator is
written in DSP itself and generates Java code.

Table \ref{experiment1} shows the performance results of 6 sample
programs executed under Windows Vista on an Intel Core2Duo 2.53 GHz
processor with 3.0 GB memory.  Java 1.6, Prolog Cafe 1.2.5, and
B-Prolog 7.4 \cite{Zhou2011a:PROLOG} were used in the experiments.
Because the Java garbage collector affects the performance, 512 MB
memory was statically allocated for the heap in all cases except for
one \footnote{About 1000 MB was allocated for the generated code for
  \Code{tarai w/o cuts}.}.

Program \Code{plan} is a simple architecture design program for a
parking structure.  It can enumerate all possible column layouts for
the given design conditions, such as free land space and the number of
stories.  Programs \Code{nqueens}, \Code{ack}, and \Code{tarai} are
well-known benchmarks, with \Code{ack} and \Code{tarai} using green
cuts for guards in Prolog, while \Code{ack w/o cuts} and \Code{tarai
  w/o cuts} do not use cuts for guards.  In the case of DSP,
\Code{ack} and \Code{tarai} use \Code{dcall} for self-recursive calls
not to leave choice points, while \Code{ack w/o cuts} and \Code{tarai
  w/o cuts} use \Code{call}.  The programs written in DSP are compiled
into Prolog and then compiled into bytecode.  The programs are forced
to backtrack in each iteration to enumerate all solutions, and the
execution times in milliseconds are averages over 10 trials.

These results show that the proposed translator generates over 25
times faster code than Prolog Cafe, over 2 times faster code than
B-Prolog, and over 5 times faster code than DSP on top of Inside
Prolog for \Code{plan} and \Code{nqueens}.  On the other hand, for
\Code{ack} and \Code{tarai} the translator generates about 2 to 3
times faster code than Prolog Cafe, but about 5 to 15 times slower
code than B-Prolog.  The translator also generates about 8 to 13 times
faster code than Prolog Cafe, but about 4 to 10 times slower code than
B-Prolog for \Code{ack w/o cuts} and \Code{tarai w/o cuts}.  Here,
\Code{plan} and \Code{nqueens} are nondeterministic, while \Code{ack}
and \Code{tarai} are deterministic.  \Code{ack w/o cuts} and
\Code{tarai w/o cuts} are also deterministic, but they involve
backtracking because of the lack of green cuts.

These experiments indicate that the proposed translation techniques
can generate faster code than Prolog Cafe and DSP on top of Inside
Prolog for all 6 programs, and faster code than B-Prolog for
nondeterministic programs.  In the case of deterministic programs, the
advantage of the proposed translation techniques is obvious against
Prolog Cafe if green cuts are not used in Prolog.  The reason why
these distinctive differences are observed seems to be that the
simplification of the variable and register management for
backtracking contributes to the performance improvement of
nondeterministic programs, but it is not effective for deterministic
programs with green cuts.

In the case of B-Prolog, the execution time of \Code{tarai} is almost
the same as that of \Code{tarai w/o cuts}.  This is because B-Prolog
compiler reduces choice points using matching trees for both
\Code{tarai} and \Code{tarai w/o cuts} \cite{Zhou1993a:PROLOG}.
Although the DSP language has no explicit cut operator of Prolog,
improving the performance by inserting cut instructions automatically
in the case of exclusive \Code{when} conditions is a future issue.

The number of instances created during an execution has a negative
impact on performance because of the garbage collection.  Obviously,
the number of instances created by the generated code for the proposed
translation techniques is greater than that for Prolog Cafe.  In the
case of \Code{tarai w/o cuts}, the generated code requires more memory
than others to prevent the garbage collection.  In the example in
Fig. \ref{PointInQuarterCircle-example1}, it is clear that the number
of instances can be reduced by merging class \Code{Method\_1\_cu1}
with class \Code{Method\_1}.  Improving the performance by the
reduction of instance creation is an important future issue.

\begin{table}
\begin{center}
\caption{Experimental results (in milliseconds)}
\begin{tabular}{l|r|r|r|r} \hline \hline
\multicolumn{1}{c|}{Program} &
\multicolumn{1}{|c|}{DSP on Prolog} &
\multicolumn{1}{|c|}{B-Prolog} &
\multicolumn{1}{|c|}{Prolog Cafe} &
\multicolumn{1}{|c}{Translator} \\ \hline
plan            &  685.0 & 295.1 & 2519.4 &  90.5 \\
nqueens         &  594.9 & 296.2 & 3279.2 & 120.3 \\
ack             & 1568.2 & 52.9 &  990.7 & 265.0 \\
tarai           & 1302.7 & 49.4 & 1680.1 & 740.8 \\
ack w/o cuts    & 2035.1 & 104.7 & 3421.3 & 403.9 \\
tarai w/o cuts  & 1307.8 &  49.2 & 6282.2 & 489.5 \\
\hline
\end{tabular}
\label{experiment1}
\end{center}
\end{table}

\section{Conclusions}\label{conclusion}

This paper described the techniques for translating the
nondeterministic functional language DSP based on attribute grammars
into Java.  The DSP is designed for knowledge representation of large
scale and complicated expert knowledge in application domains.  It is
capable of representing trial and error without any side-effects or
loop constructs using nondeterministic features.  Current development
and runtime environments are built on top of Inside Prolog, while the
runtime environment can be embedded in a Java-based application
server.  However, issues regarding language interoperability and
adaptability to new computer environments are envisaged when applied
to practical application development.  The language translation is
intended to improve the interoperability and adaptability of DSP.

The proposed translation techniques are based on binarization and the
techniques proposed for the translation of Prolog.  The performance,
however, is improved by introducing the continuation unit and
simplifying the management of variables and registers using the
semantic differences of variables and explicit determinism of some
statements.  An experimental translator written in DSP itself
generates Java code from DSP descriptions, and the experimental
results indicate that the generated code is over 25 times faster than
that of Prolog Cafe for nondeterministic programs, and over 2 times
faster for deterministic programs.  The generated code is also over 2
times faster than B-Prolog for nondeterministic programs.  However,
the generated code is about 3 to 15 times slower than B-Prolog for
deterministic programs.  Improving the performance of deterministic
programs is an important future issue.


\end{document}